\documentclass[preprint2]{aastex}

\slugcomment{}

\shorttitle{Cherenkov-type radiation in pulsars}
\shortauthors{Machabeli /& Chkheidze}

\begin{document}


\title{On high frequency Cherenkov-type radiation in pulsar magnetospheric electron-positron
plasma}


\author{G. Machabeli and N. Chkheidze}
\affil{Centre for Theoretical Astrophysics, ITP, Ilia State
University, 0162-Tbilisi, Georgia}
    \email{g.machabeli@iliauni.edu.ge}


\begin{abstract}
Emission process of a charged particle propagating in a medium with
a curved magnetic field is considered. This mechanism combines
features of conventional Cherenkov and curvature emission. Thus,
presence of a medium with the index of refraction larger than the
unity is essential for the emission. In the present paper the
generation of high frequency radiation by the mentioned mechanism is
considered. The generated waves are vacuum-like electromagnetic
waves and may leave the medium directly. Consequently, this emission
mechanism may be important for the problem of pulsar X-ray and
gamma-ray emission generation.
\end{abstract}

\keywords{pulsars: general --- radiation processes: nonthermal}

\section{Introduction}

In the electron-positron ($e^{-}e^{+}$) plasma filling the
magnetosphere of pulsars can generate waves in the different
frequency ranges. It is well known, that relativistic charged
particle moving along the curved magnetic field lines emit
radiation. In this context was considered the analogy of curvature
radiation in the pulsar magnetosphere, for the generation of radio
waves. It was found that for the beam of particles the wave
interfence processes can suppress the generated emission
\citep{bla,me}. Later on the similar processes were reconsidered,
taking into account the inhomogeneity, particularly the curvature of
the field lines, which causes the drift motion of the particles
\citep{ka,ly}. It appeared that the $e^{-}e^{+}$ plasma can generate
radio emission in this case, through the modified
Cherenkov-curvature resonance
\begin{equation}\label{}
    \omega-k_{\parallel}\upsilon_{\parallel}-k_{\perp}u_{\perp}=0,
\end{equation}
where, signs $\perp$ and $\parallel$ denote components across and
along the pulsar magnetic field $B_{0}$,
$u_{\perp}=\upsilon_{\parallel}^{2}\gamma_{r}/R_{c}\omega_{B}$ is
the drift velocity of the particles due to curvature of the magnetic
field lines ($R_{c}$ is the curvature of the field lines,
$\omega_{B}=eB/mc$, $B$ is the pulsar magnetic field, $e$ is the
electrons charge and $m$ is its rest mass, $c$ is the speed of
light). The curved field lines lie on a plane across which the
particles perform the drift motion. This particularly violates the
axial symmetry of the mentioned problem, leading to coupling of the
parallel and perpendicular components of electric field of waves,
provoking appearance of additional terms in the dispersion relation,
which maintain the resonance condition (1). The issue of generation
of electromagnetic radio waves in approximation of infinitely large
magnetic field $B\rightarrow\infty$ was thoroughly investigated in
the following works \citep{ge1,ge2}. In these papers the medium
inhomogeneity is neglected and one considers the oblique propagation
of waves, which results the violation of the axial symmetry, as
well. Even choosing the wave vector in the following way
$\textbf{k}=(k_{\perp},0,k_{\parallel})$ does not help to separate
the parallel and perpendicular components of electric field of waves
$E_{\parallel}$ and $E_{\perp}$, in the dispersion relation. Which
by itself leads to terms containing the Cherenkov resonance in the
dispersion relation
\begin{equation}\label{}
    \omega-k_{\parallel}\upsilon_{\parallel}=0.
\end{equation}
The fact that this is a Cherenkov-type resonance immediately implies
that the presence of a subluminous wave with the phase velocity
smaller than the speed of light is essential. It is well known, that
if the speed of charged particle is greater than the speed of light
in the medium the Cherenkov radiation is generated \citep{che}.

We suppose that the Cherenkov-curvature resonance can provide
generation of high frequency (X-ray and gamma-ray) radiation, as
well. This mechanism has been so far considered only for explanation
of pulsar radio emission generation in the pulsars' magnetosphere.
Differently from radio emission, the high frequency radiation does
not 'feel' the medium and emerges freely from the region of its
generation, as in the pulsar magnetosphere for X-ray and higher
energy emission the condition $\lambda\ll n^{-1/3}$ is satisfied
(here $n$ is the density of plasma particles, and $\lambda$ is the
wave-length of high frequency waves). Consequently, the emission of
single particles can be summed up without taking into account the
interference of waves. Generally accepted generation mechanisms of
the high energy emission in pulsar magnetospheres are the
synchrotron-curvature radiation and the inverse-Compton scattering.
The Cherenkov type radiation has not been so far considered, as
possible mechanism for generation of high energy emission in pulsar
magnetospheres. In the present paper we investigate generation of
high frequency ($\omega\gg\omega_{B}/\gamma_{b}$) waves in the
relativistic $e^{-}e^{+}$ plasma filling the pulsar magnetosphere
via the Cherenkov-curvature resonance.

\section{Linear Waves in Pair ($e^{-}e^{+}$) Plasma}

\begin{figure}[t]
\includegraphics[width=6.5cm]{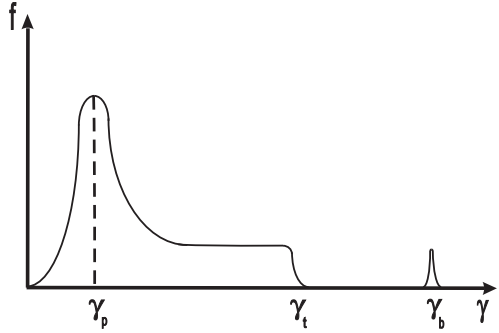}
\caption{\label{fig:1} The distribution function of a
one-dimensional plasma in the pulsar magnetosphere. Left
correspoends to the secondary particles, right to the primary beam}
\end{figure}

\begin{figure}[t]
\includegraphics[width=7cm]{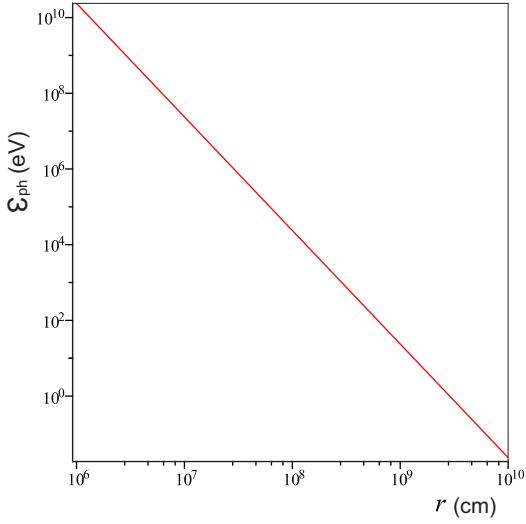}
\caption{\label{fig:2} The dependence of maximum possible energy of
photon $\varepsilon_{ph}$ radiated through the Cherenkov mechanism
in the pulsar magnetosphere on the distance from the star surface
$r$}
\end{figure}

\begin{figure}[b]
\includegraphics[width=7cm]{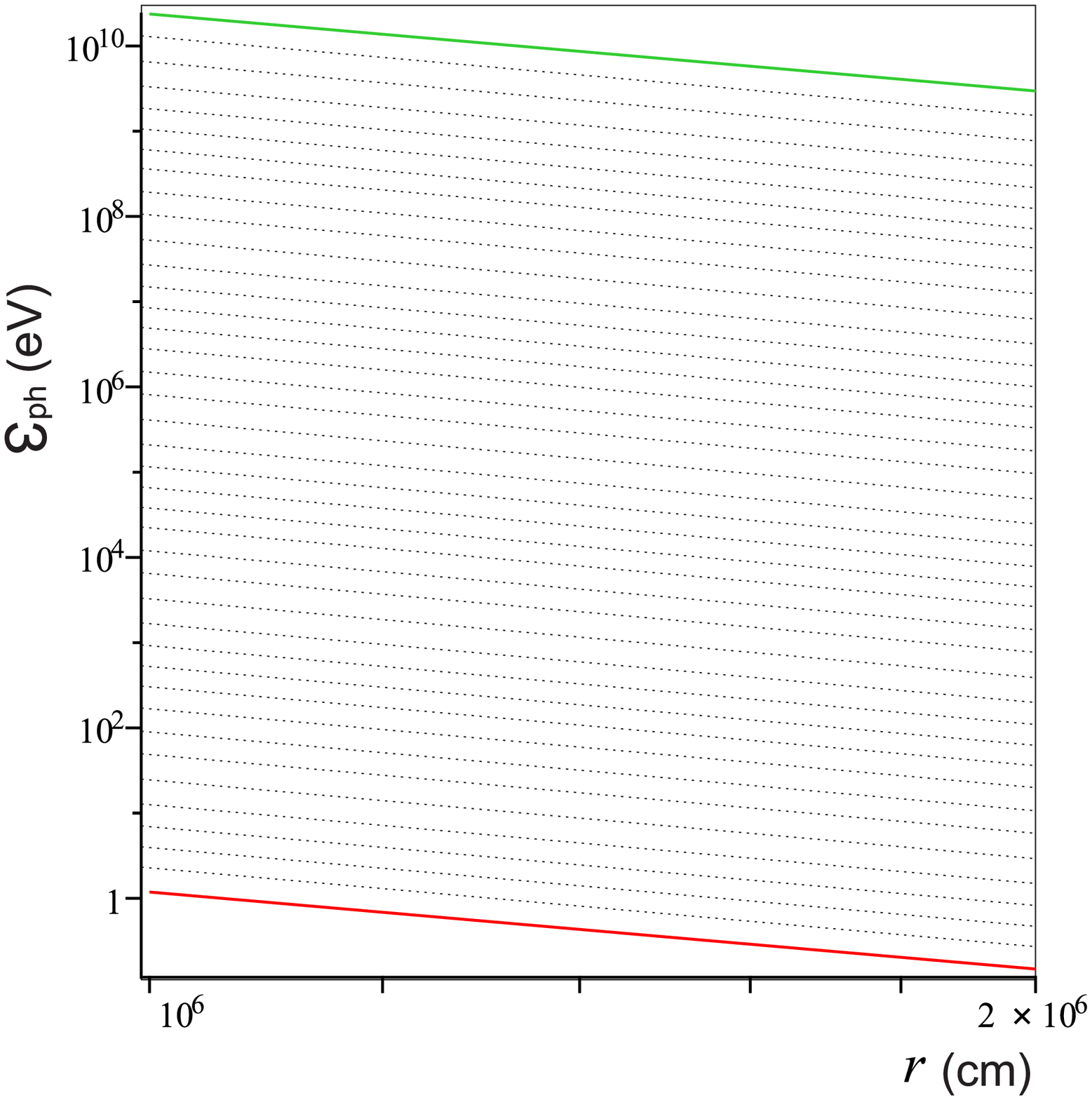}
\caption{\label{fig:3} The energy range of photons radiated through
the Cherenkov mechanism in the pulsar magnetosphere near the star
surface}
\end{figure}

Let us consider the wave propagation in $e^{-}e^{+}$ plasma of
pulsar magnetosphere with small inclination angles to the magnetic
field $B_{0}$, when the waves electric field components
$E_{\parallel}$ and $E_{\perp}$ are bounded with each other. In this
case for the electromagnetic waves, the Maxwell's equations mutually
with the Vlasov equation for the pair plasma in the linear
approximation can be written as \citep{vo}:
\begin{eqnarray}\label{}
     \left(k_{\parallel}^{2}c^{2}-\omega^{2}\varepsilon_{xx}\right)\left(k_{x}^{2}c^{2}-\omega^{2}\varepsilon_{zz}\right)=\nonumber \\
     =\left(k_{x}k_{\parallel}c^{2}-\omega^{2}\varepsilon_{xz}\right)^{2}.
\end{eqnarray}
Here $\omega$ is the wave frequency, $\varepsilon_{ij}$ is the
dielectric tensor for plasma ($i,j=x,y,z$) and $B_{0}\parallel z$.
The components of dielectric tensor written for the weakly curved
magnetic field lines ($u_{\perp}/c\ll1$) have the following form
\citep{ka,ly}
\begin{eqnarray}\label{}
  \varepsilon_{xx}=1-\frac{1}{2}\Sigma_{a}\frac{\omega_{pa}^{2}}{\omega^{2}}\int\frac{dp_{\parallel}}{\gamma}(\omega-k_{\parallel}v_{\parallel})A^{+}_{a}f_{a}-
  \nonumber
  \\
  -\Sigma_{a}\omega_{pa}^{2}\int\frac{dp_{\parallel}}{\gamma}\frac{f_{a}}{\Omega_{a}^{0^{2}}}\frac{k_{\parallel}^{2}u_{a}^{2}}{\omega^{2}}\left(1-\frac{\omega v_{\parallel}}{k_{\parallel}c^{2}}\right),\nonumber
 \\
  \varepsilon_{yy}=1-\frac{1}{2}\Sigma_{a}\frac{\omega_{pa}^{2}}{\omega^{2}}\int\frac{dp_{\parallel}}{\gamma}(\omega-k_{\parallel}v_{\parallel}-k_{x}u_{a})A^{+}_{a}f_{a}
  \nonumber
   \end{eqnarray}
  \begin{eqnarray}\label{}
  \varepsilon_{zz}=1-\Sigma_{a}\omega_{pa}^{2}\int\frac{dp_{\parallel}}{\gamma}\frac{f_{a}}{\Omega_{a}^{0^{2}}}\left[1-\frac{k_{x}u_{a}}{\omega}\right]\times
  \nonumber
  \\  \times
  \left[\left(1-\frac{k_{x}u_{a}}{\omega}\right)-\frac{v_{\parallel}^{2}}{c^{2}}\right], \nonumber
  \\
  \varepsilon_{xz}=-\varepsilon_{zx}=-\frac{1}{2}\Sigma_{a}\frac{\omega_{pa}^{2}}{\omega^{2}}\int\frac{dp_{\parallel}}{\gamma}k_{x}
  v_{\parallel}- \nonumber \\
  -\Sigma_{a}\omega_{pa}^{2}\int\frac{dp_{\parallel}}{\gamma}\frac{f_{a}}{\Omega_{a}^{0^{2}}}\frac{k_{\parallel}u_{a}}{\omega},
\end{eqnarray}
where
\begin{eqnarray}\label{}
    A^{+}_{a}=\left(\frac{1}{\Omega_{a}^{+}}+\frac{1}{\Omega_{a}^{-}}\right),
\end{eqnarray}
\begin{equation}\label{}
    \Omega_{a}^{\pm}=(\omega-k_{\parallel}v_{\parallel}-k_{x}u_{a}\pm
    \omega_{B}/\gamma), \qquad \Omega_{a}^{0}=(\omega-k_{\parallel}v_{\parallel}-k_{x}u_{a}).
\end{equation}
Here $\omega_{pa}^{2}=4\pi e^{2}n_{pa}/m$, $n_{pa}$ is the
concentration of a-type particles, $f_{a}$ is the one-dimensional
distribution function of a components of $e^{-}e^{+}$ plasma. Due to
specific mechanism of filling the pulsar magnetosphere with charged
particles, the distribution function of pair $e^{-}e^{+}$ plasma
consists of three components: the bulk of plasma with the
Lorentz-factor $\gamma_{p}\approx(3-10)$ and the particle
concentration at the star surface $n_{p}\approx10^{20}$cm$^{-3}$,
the long extended in one direction tail on the distribution function
with $\gamma_{t}\approx(10^{4}-10^{5})$,
$n_{t}\approx(10^{16}-10^{17})$cm$^{-3}$ and the most energetic
primary beam with $\gamma_{b}\approx(10^{6}-10^{7})$ \citep{os},
$n_{b}\approx(10^{13}-10^{14})$cm$^{-3}$ \citep{gold,stur,tad,ru}
(see Fig.~\ref{fig:1}). In the expression for the dielectric tensor
$\varepsilon_{ij}$, when summing up by different types of particles
one should only take into account the contribution of bulk plasma.
The situation changes, when the resonance conditions are satisfied
for the beam particles (Exp. (1)). As in this case the beam
particles appear to have the equal contribution in expression (4),
as the bulk particles. Consequently, in the hydrodynamical
approximation for the particle distribution function we apply
summing for beam and bulk plasma particles $f_{a}=\delta (v-v_{a})$,
$a=p,b$.

Let us consider the possibility of generation of high frequency
waves, which satisfy the following conditions
$\omega_{B}^{2}/\gamma_{b}^{2}\ll1\ll\omega_{B}^{2}/\gamma_{p}^{2}$.
The frequency of the excited waves can be written as
$\omega=kc(1-\Delta)$, where $\Delta\ll1$ and
$k=k_{\parallel}(1+\tan^{2}\theta)^{1/2}$. Here $\tan
\theta=k_{\perp}/k_{\parallel}$ and as we are considering the small
angles of propagation ($\theta\ll1$), one can write $k\approx
k_{\parallel}\left(1+\theta^{2}/2\right)$. We assume that $k_{y}=0$
and taking into account the following conditions $u_{d}/c\ll1$,
$\omega\gg\omega_{p}\gamma_{p}$, $\omega\gg\omega_{B}/\gamma_{b}$
and $1/\gamma_{b}^{2}\ll\Delta\ll1/\gamma_{p}^{2}$. The components
of dielectric tensor calculated from Eq. (4) take the form:
\begin{eqnarray}\label{}
    \varepsilon_{xx}=\varepsilon_{yy}=1, \nonumber \\
    \varepsilon_{zz}=1+\frac{\omega_{b}^{2}}{\gamma_{b}}\frac{1}{k_{\parallel}^{2}c^{2}\Delta^{2}}\theta\frac{u}{c},
    \nonumber \\
    \varepsilon_{xz}=\frac{\omega_{b}^{2}}{\omega^{2}}\frac{\theta}{\gamma_{b}}\frac{\Delta}{\Delta^{2}-\frac{\omega_{B}^{2}}{k_{\parallel}^{2}c^{2}\gamma_{b}^{2}}}.
\end{eqnarray}
It should be mentioned that the above expressions were taken,
assuming that $\theta/2=u/c$. Using this expressions for
$\varepsilon_{ij}$, in place of Eq. (3) we will get:
\begin{equation}\label{}
    \Delta^{2}-\frac{\omega_{b}^{2}}{2k_{\parallel}^{2}c^{2}\gamma_{b}}\Delta-\frac{\omega_{B}^{2}}{k_{\parallel}^{2}c^{2}\gamma_{b}^{2}}=0.
\end{equation}
From where:
\begin{equation}\label{}
    \Delta=\frac{\omega_{b}^{2}}{4k_{\parallel}^{2}c^{2}\gamma_{b}}\pm \left[\left(\frac{\omega_{b}^{2}}{4k_{\parallel}^{2}c^{2}\gamma_{b}}\right)^{2}+\frac{\omega_{B}^{2}}{4k_{\parallel}^{2}c^{2}\gamma_{b}^{2}}\right]^{1/2}.
\end{equation}
Taking into account the estimations done for the parameters, this
expression can be reduced to the following form:
\begin{equation}\label{}
    \Delta=\frac{\omega_{B}}{k_{\parallel} c \gamma_{b}}.
\end{equation}
From the general equations for fields \citep{vo,aro}, its easy to
define the polarization of this waves when $\theta\neq 0$,
$E_{x}/E_{z}\gg1$ and the wave is practically linearly polarized,
its electric field vector lies in the $\mathbf{k},\mathbf{B_{0}}$
plane. The phase velocity of these waves can be defined as
\begin{equation}\label{}
    v_{ph}=\frac{\omega}{k}=c(1-\Delta).
\end{equation}
The value of $\Delta$ defines how the phase velocity of the wave in
the $e^{-}e^{+}$ differs from the speed of light in the vacuum. The
condition
\begin{equation}\label{}
    v_{b}-v_{ph}>0,
\end{equation}
defines the generation of Cherenkov emission. We write the
relativistic velocity of the particles in the following way
$v_{b}\approx c(1-1/(2\gamma_{b}^{2}))$ and in this case the
condition rewrites as
\begin{equation}\label{}
    \Delta>\frac{1}{2\gamma_{b}^{2}}.
\end{equation}
The condition (13) and expression (10) defines the conditions for
generation of hight energy Cherenkov emission in the pulsar
magnetosphere
\begin{equation}\label{}
    k_{\parallel}c<2\gamma_{b}\omega_{B_{0}}\left(\frac{r_{0}}{r}\right)^{3},
\end{equation}
where $\omega_{0}=eB_{0}/mc$ is the cyclotron frequency at the
surface and $r_{0}$ is the star radius. Let us assume that the
distance changes from star surface $r=r_{0}$ to the light cylinder
radius $r=r_{LC}$, which can be taken as the limit distance for the
generation region of pulsed emission in pulsars. Then one can find
from Eq. (14) the maximum photon energy that can be radiated through
the Cherenkov mechanism at the certain distance
\begin{equation}\label{}
    \varepsilon_{ph}\approx8.3\cdot10^{-15}\gamma_{b}\omega_{B_{0}}\left(\frac{r_{0}}{r}\right)^{3}.
\end{equation}
Assuming the typical parameters for radio pulsars,
$B_{0}\simeq10^{12}$G, $\gamma_{b}\simeq10^{6}$ and the neutron star
radius $r_{0}\simeq10^{6}$cm, we find that the most energetic
photons can be emitted near the star surface (see Fig.~\ref{fig:2}).
In particular, at the star surface the beam electrons emitting
through the Cherenkov mechanism can generate high energy gamma-ray
photons ($\varepsilon_{ph}\simeq24$GeV). On the other hand emitting
through the same mechanism at the light cylinder distances
($r_{LC}\sim10^{8}$cm) the radiation comes in X-ray domain
($\varepsilon_{ph}\simeq24$KeV). As we are searching for the
generation of high frequency emission by Cherenkov mechanism, more
interesting is to consider the process near the star surface, where
the upper limit on generated radiation comes in gamma-ray domain. It
is important to find also the lower limit of Cherenkov emission
generated near the star surface. For this purposes, we take into
account the conditions for the wave frequency that were assumed
during the calculations. In particular, the highest lower limit for
the wave frequency can be obtained from the following condition
\begin{equation}\label{}
    \Delta\ll\frac{1}{\gamma_{p}^{2}}.
\end{equation}
Taking into account the Eq. (10) for the $\Delta$, one obtains
\begin{equation}\label{}
    k_{\parallel}c\gg\frac{\omega_{B_{0}}\gamma_{p}^{2}}{\gamma_{b}}\left(\frac{r_{0}}{r}\right)^{3}.
\end{equation}
Using expression (14) and (17) we define the energy domain for the
Cherenkov radiation generated near the star surface, which covers
X-ray up to gamma-ray domains (see Fig.~\ref{fig:3}).

It is also interesting to estimate the luminosity of the high
frequency radiation generated through the Cherenkov-curvature
mechanism in the pulsar magnetosphere. The luminosity can be
calculated by the following expression
\begin{equation}\label{}
    L=4\pi r^{2} F \textit{f}_{\Omega},
\end{equation}
where $r$ shows the location of emission generation region, $F$ is
the emission flux and $\textit{f}_{\Omega}$ is the beaming fraction.
For typical radio pulsars $0.4<\textit{f}_{\Omega}<1$ \citep{nara}.
For estimations we set the value of beaming fraction to 1 and
consider the generation of the emission near the star surface
($r=r_{0}$). The emission is generated due to the kinetic energy of
the resonant particles. Consequently, for the emission flux one can
write
\begin{equation}\label{}
    F=mc^{3}\gamma_{b}n_{b},
\end{equation}
here $n_{b}$ is the density of emitting particles. Using the values
for typical radio pulsars, we find that the luminosity of the high
frequency radiation generated near the star surface
$L\sim10^{35}-10^{36}$erg/s.

\section{Conclusion}

In this paper we considered a Cherenkov-curvature emission mechanism
which combines features of conventional Cherenkov and curvature
radiation. It is essential that even a weak inhomogeneity of the
magnetic fields results in a drift motion of the particle
perpendicular to the local plane of the magnetic field line, which
is weakly relativistic when the motion of the particles along the
magnetic field is ultrarelativistic. This causes generation of
vacuum-like waves, propagating freely in the pulsar magnetosphere
that can reach a observer as pulsar radiation. Physical origin of
the emission in the case of Cherenkov-type and synchrotron-type
processes is quite different. Cherenkov-type process the emission
may be attributed to the electromagnetic polarization shock front
that develops in a dielectric medium due to the passage of a charged
particle with speed larger than phase speed of waves in a medium. It
is virtually a collective emission process. Cherenkov-type emission
is impossible in vacuum and in a medium with the refractive index
smaller than unity. The conventional Cherenkov and curvature
emission mechanisms may be viewed as corresponding limits of the
Cherenkov-curvature mechanism in the cases of homogeneous magnetic
field (in a medium), medium without magnetic field, and
inhomogeneous magnetic field without a medium. The
Cherenkov-curvature instability develops on the rising part of the
beam distribution function (see Fig.~\ref{fig:1}). The free energy
for the growth of the instability comes from the non-equilibrium,
anisotropic distribution of the fast particles. For the development
of the instability its essential that the medium supports
subluminous waves, i.e. its index of refraction is larger than
unity.

In our case, we have studied the possibility of generation of high
frequency radiation through the mentioned Cherenkov-curvature
mechanism. It revealed that this mechanism can provide excitation of
emission from the broad energy domain. In particular near the star
surface generation of X-rays up to high energy gamma-rays is
possible, if the resonant particles are the most energetic primary
beam electrons. This emission can freely emerge in the pulsar
magnetosphere and reach an observer. The interesting feature of this
type of radiation should be its angular distribution. The
calculations showed that the emission is generated in the cone
centered at the angle $\theta/2=u/c$. Consequently, one can find the
angular distribution of the generated radiation
($\theta\approx2c^{2}\gamma_{b}/R_{c}\omega_{B}$) for the chosen
location of the generation region. Taking into account that the beam
has a very narrow distribution, one should expect also a small
opening angle of the radiation cone, that inevitably will cause
detection of the narrow pulses.

\acknowledgments

The research of the authors was supported by the Shota Rustaveli
National Science Foundation grant (N31/49).

\end{document}